\documentclass[aip,cha]{revtex4-1}

\usepackage{graphicx}%
\usepackage{dcolumn}%
\usepackage{bm}%

\usepackage{amssymb}

\begin{document}

\title{On the non-randomness of maximum Lempel Ziv complexity sequences of finite size}

\author{E. \surname{Estevez-Rams}}
\email{estevez@imre.oc.uh.cu}
\affiliation{Instituto de Ciencias y Tecnolog\'ia de Materiales, University of Havana (IMRE), San Lazaro y L. CP 10400.
La Habana. Cuba.}

\author{R. \surname{Lora Serrano}}
\affiliation{Universidade Federal de Uberlandia, AV. Joao Naves de Avila, 2121- Campus Santa Monica, CEP 38408-144,
Minas Gerais, Brasil. }

\author{B. \surname{Arag\'on Fern\'andez}}
\author{I. \surname{Brito Reyes}}
\affiliation{Universidad de las Ciencias Inform\'aticas (UCI), Carretera a San Antonio, Boyeros. La Habana. Cuba.}

\begin{abstract}
Random sequences attain the highest entropy rate. The estimation of entropy rate for an ergodic source can be done using the Lempel Ziv complexity measure yet, the exact entropy rate value is only reached in the infinite limit. We prove that typical random sequences of finite length fall short of the maximum Lempel-Ziv complexity, contrary to common belief. We discuss that, for a finite length, maximum Lempel-Ziv sequences can be built from a well defined generating algorithm, which makes them of low Kolmogorov-Chaitin complexity, quite the opposite to randomness. It will be discussed that Lempel-Ziv measure is, in this sense, less general than Kolmogorov-Chaitin complexity, as it can be fooled by an intelligent enough agent. The latter will be shown to be the case for the binary expansion of certain irrational numbers.  Maximum Lempel-Ziv sequences induce a normalization that gives good estimates of entropy rate for several sources, while keeping bounded values for all sequence length, making it an alternative to other normalization schemes in use.
\end{abstract}

\date{\today}

\maketitle

\begin{quotation}
Lempel and Ziv showed that measuring the capacity of an ergodic source to generate new patterns could be used to estimate its entropy rate. Entropy rate, is a length invariant measure of the amount of new information gained per unit step in a dynamical process. It has been found ubiquitous in a number of areas, such as the study of dynamical systems, information theory (from where its definition is usually drawn), or complexity theory; and it has become one of the fundamental concepts in data analysis. Lempel-Ziv complexity (LZ76) method of estimating the entropy rate from a single discrete series of measurement, carries a number of practical advantages. It is generally assumed that random sequences, being of maximum entropy rate, attain maximum Lempel-Ziv complexity for sequence of finite length. We prove that, contrary to this common belief, finite random sequences do not attain maximum Lempel-Ziv complexity. The consequence is that normalization of LZ76 complexity by that of a random sequence, does not yield a properly bounded value, and estimated entropy rates in this way derived, can achieved inconsistent values above one. Instead, there is perfectly deterministic way, and therefore non-random procedure, of generating maximum Lempel-Ziv sequences (MLZs) that can be used for normalization purposes, yielding sound estimates of entropy rate. 
\end{quotation}

\section{Introduction}

Estimating entropy rate from a single finite measurement is far from being a simple task. Lempel-Ziv complexity measure (from now on LZ76 complexity), as described in the seminal paper of Lempel and Ziv\cite{lz76}, stems its popularity from the fact that it is straight forward to calculate for experimental data\cite{amigo03,szczepanski04,abo06,hu06,zhang09} and, to the essential property, that its growth rate is closely related to the entropy rate for an ergodic source\cite{ziv78}. LZ76 complexity has been studied from different points of view, and used in different context (see for example Ref. \onlinecite{chelani11,rajkovic03,liu12}). 

LZ76 complexity is considered as a model based, non-probabilistic, randomness finding measure of complexity\cite{rapp01}. Randomness finding is used for those measures that give the highest values of complexity to random sequences, being Kolmogorov-Chaitin complexity (KC complexity) the most encompassing complexity measure of this kind\cite{kolmogorov65}.  KC complexity measures the length of the shortest program, run in a Universal Turing Machine, that allows to reproduce the analyzed sequence. It is closely related to randomness in the sense of those infinite sequences which pass the Universal Martin-L\"of test of randomness\cite{martin66}: All Martin-L\"of random sequences have maximum KC complexity. Those ideas have been extended to the case of finite length sequences\cite{li94}. By definition, KC complexity can not be outsmarted by an intelligent agent (or any other agent for that matter). Any ''thought of'', clever sequence, has, necessarily, a shorter algorithmically description than the length of the sequence and therefore, a smaller KC complexity than a random string of the same length. Unfortunately, it is impossible to find, in a systematic way, the smaller program reproducing a given sequence or, to assert if a given program is the smallest possible\cite{calude02}.

If we consider the set of all strings of length N, it is often assumed that LZ76 complexity attains its maximum value for those finite strings for which the KC complexity is maximum, that is, for all sequences that behaves random enough\cite{li94}. Yet, this assumption is wrong. In this contribution we will explore the nature of maximum LZ76 sequences (we will denote them by MLZs) as built from an algorithm much shorter than the
sequence length itself. It will be shown that the MLZs have, for a finite length N, larger LZ76 complexity than the typical\cite{Note1} random string of the same length. The very algorithmic nature of the MLZs make them the opposite to randomness in the KC complexity sense. The implications of such analysis, is that LZ76 complexity can indeed be outsmarted by and intelligent agent (or a sophisticated enough automata) contrary to the KC complexity. LZ76 complexity measures a deeper property of sequences than that of randomness alone.

Having devised a procedure to build MLZs we turn to the question of using them as an effective normalization factor for LZ76 complexity. Normalization of LZ76 complexity is of concern when comparison of different complexity measures is desired. This issue has been discussed before\cite{zhang09,rapp01,lesne09}. It will be convenient to have a normalized complexity measure that lies between well defined values i.e. in the interval $[0,1]$. Most importantly, normalization should be compatible with Ziv theorem that, in the infinite limit, the normalized LZ76 complexity will tend to the entropy rate for an ergodic source\cite{ziv78}.  Finally, the results will be applied to known entropy dynamics such as, the logistic map, the biased Bernoulli process and finite state automata.

\section{Lempel-Ziv complexity}\label{sec:lz76}

Consider the following factorization of a  sequence $u=u_1u_2\dots u_N$: 

\[E(u)=u(1,h_1)u(h_1+1,h2)\dots u(h_{m-1}+1,N), \]

where $u(i,j)$ is the substring $u_iu_{i+1}\dots u_j$, and each symbol $u_i$ is drawn from a finite alphabet $\Sigma$ of cardinality $\sigma(=|\Sigma|)$. $E(u)$ is called an exhaustive history of the sequence $u$, if any factor $u(h_{j-1}+1, h_j)$ is not a substring of the string $u(1,h_j-1)$, while $u(h_{j-1}+1, h_{j}-1)$ is. The LZ76 complexity $C(u)$, is then the cardinality (number of factors) of the exhaustive history $E(u)$. For example, the exhaustive history of the sequence $u=010011101101100$ is $E(u)=0.1.00.11.101.101100$, where a dot is used to separate each factor, and $C(u)=6$.

In general, $C(u)$ for a length $N$ string, is bounded by\cite{lz76}

\begin{equation}\label{ineq}
 C(u) < \frac{N}{(1-\varepsilon_N)\log_{\sigma}{N}},
\end{equation}

where

\begin{equation}\label{epsilon}
 \varepsilon_{N}=2\frac{1+\log_{\sigma}{\log_{\sigma} \sigma N}}{\log_{\sigma}N}.
\end{equation}

In what follows, we will use $\log{x}\equiv \log_{\sigma}{x}$ to simplify the notation. $\varepsilon_{N}$ is a slowly decaying function of $N$, leading to an asymptotic value

\begin{equation}\label{asymp}
 C(u) < \frac{N}{\log{N}},
\end{equation}

for large enough $N$.

Ziv\cite{ziv78} proved that, if $u$ is the infinite length output from an ergodic source with entropy rate $h$, then

\begin{equation}\label{zivtheorem}
\limsup_{N\rightarrow\infty}\frac{C[u(1,N)]}{N/\log{N}}=h.
\end{equation}

almost surely\cite{Notereferee}. Here the entropy rate\cite{Note2} is taken by its information theoretical definition

\begin{equation}\label{shannonh}
 h= \lim_{N\rightarrow\infty} h(N)=\lim_{N\rightarrow\infty}\frac{H(N)}{N}
\end{equation}

where $H(N)$ is the Shannon block entropy\cite{cover06} over the length $N$ substrings $u(j,j+N-1)\in u$.

Equation (\ref{zivtheorem}) induces to define a normalized LZ76 complexity by

\begin{equation}\label{lz76c}
 c[u(1,N)]=\frac{C[u(1,N)]}{N/\log{N}}
\end{equation}

Care must be taken in the use of equation (\ref{ineq}). Strictly speaking, inequality (\ref{ineq}) is valid as long as $\varepsilon_{N} < 1$, as $C(u)$ is positive defined. This point is not clarified in the original deduction\cite{lz76}. For $\varepsilon_{N}\sim 1$, the upper bound (\ref{ineq}) simply diverges making it of little use. For alphabet size of $\sigma=27$ symbols, used in the complexity analysis of English texts\cite{schurmann99}, this lower limit for the string size $N$, is already above $10^4$ symbols. Even more significant, is the very slow decreasing nature of $\varepsilon_N$ which, for a binary alphabet ($\sigma=2$) reaches $\varepsilon_N=0.1$ for $N=4\times 10^{50}$. Equation (\ref{asymp}) is only valid for very large values of $N$, usually not attainable in real experimental data. The conclusion is that, although for an ergodic source equation (\ref{zivtheorem}) is valid  in the limit of infinite length sequences, $c(u(1,N))$ can have values above $1$, even for very long, yet finite, strings. It will be wrong to identify $h[u(1,N)]$, with $c[u(1,N)]$, for one, the former is always less than $1$, which is not the case for the latter. It is also important to clarify that LZ76 complexity as defined by equation (\ref{lz76c}) and introduced in Ref. \onlinecite{lz76}, must not be confused with entropy rate estimation based on the length of an encoded string using some encoding procedure as, for example, those derived from the Lempel-Ziv algorithm\cite{lz77} or related compression schemes\cite{Note4}.  Finally, it must be pointed out that, from the slow convergence of $\varepsilon_N$, estimating entropy from LZ76 complexity must be made with special care as it has been studied in an number of reports\cite{rapp01,lesne09}.

Equation (\ref{zivtheorem}) infers that random sequences, as those output from a $(\frac{1}{2},\frac{1}{2})$-Bernoulli process (fair coin toss), reach the highest possible $c(u)$ value for an infinite length. Yet, equation (\ref{zivtheorem}) does not imply that this bound value is only attainable by a random sequence.

From the very algorithmic definition of LZ76 complexity, it is clear that a finite procedure can be devised, of size at least $O(\log{N})$, that builds a maximum Lempel-Ziv complexity sequence in a deterministic way with slow increasing KC complexity.

\section{Maximum Lempel-Ziv complexity sequences}\label{sec:mlzs}

For a given length $k$, there will be $\sigma^k$ different strings. Consider the following generating process. First, output the $\sigma$ characters of the alphabet in lexicographic order. Each character will contribute as a component to the exhaustive history. Next, consider the set of all strings of length two ordered in lexicographic order. Output a string of this set, if, and only if, contributes as a component to the exhaustive history. Once all
strings of length two are considered, the process is repeated for strings of length three, four and so on. The resulting output string will have, by construction, the maximum Lempel-Ziv complexity among all strings of the same length. As an example, we show a maximum LZ76 complexity string of length $39$, for a binary alphabet

$0.1.00.11.000.101.0000.0111.1011.00100.01011.01110.$

where a dot is used to separate each factor in the exhaustive history.

For a given length and alphabet size, the maximum LZ76 complexity string is not unique. In the above algorithm, there is no need to test the candidate components of length $l$ in lexicographic order. One can take any ordering for choosing the factor being tested, including random ordering. Numerical simulations show little dependence of the LZ76 complexity of the MLZs on the ordering of the test step.

\begin{figure}
\includegraphics{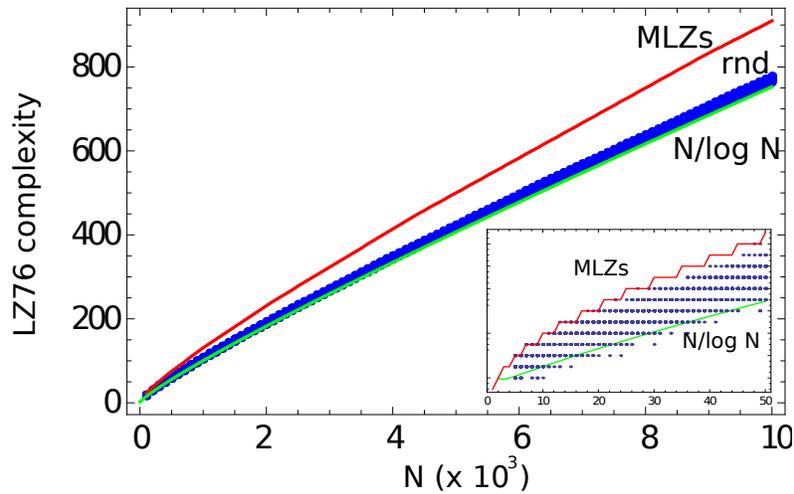}
\caption{LZ76 complexity as a function of sequence length $N$, for MLZs and for $10^2$ random sequences. The MLZs upper bound is clearly observed, while the simulated  random sequences (rnd) are below the MLZs values and mostly above the $N/\log{N}$ curve. In the inset it can be seen that LZ76 complexity for the random sequences can lie also below the $N/\log{N}$ curve.
}\label{mlzsrandom}
\end{figure}

Figure \ref{mlzsrandom} plots LZ76 complexity as a function of sequence length for the MLZs, together with the plot of $N/\log{N}$, and that of $10^2$ random binary strings. While the MLZs are clearly an upper bound for the LZ76 complexity, that is not the case for the right hand of equation (\ref{asymp}), a fact further emphasized in the inset of the same figure for smaller $N$ values. Random strings fail on average to reach the maximum LZ76 complexity for almost all length considered. For small values of $N$, random sequences do indeed fill the space between
$N/\log{N}$ and MLZs (see the inset in figure \ref{mlzsrandom}), but as the length increases, it starts in average to fall increasingly below the LZ76 complexity value of the MLZs at least up to $N=10^6$. Asymptotically, it seems that the slope of the LZ76 complexity curve for the MLZs starts decreasing and eventually merges with that of the random sequence and of the $N/\log{N}$ curve, yet, as already discussed, this can happen for very large values ($>10^{50}$) of sequence lengths.

\begin{figure}
\includegraphics[scale=0.8]{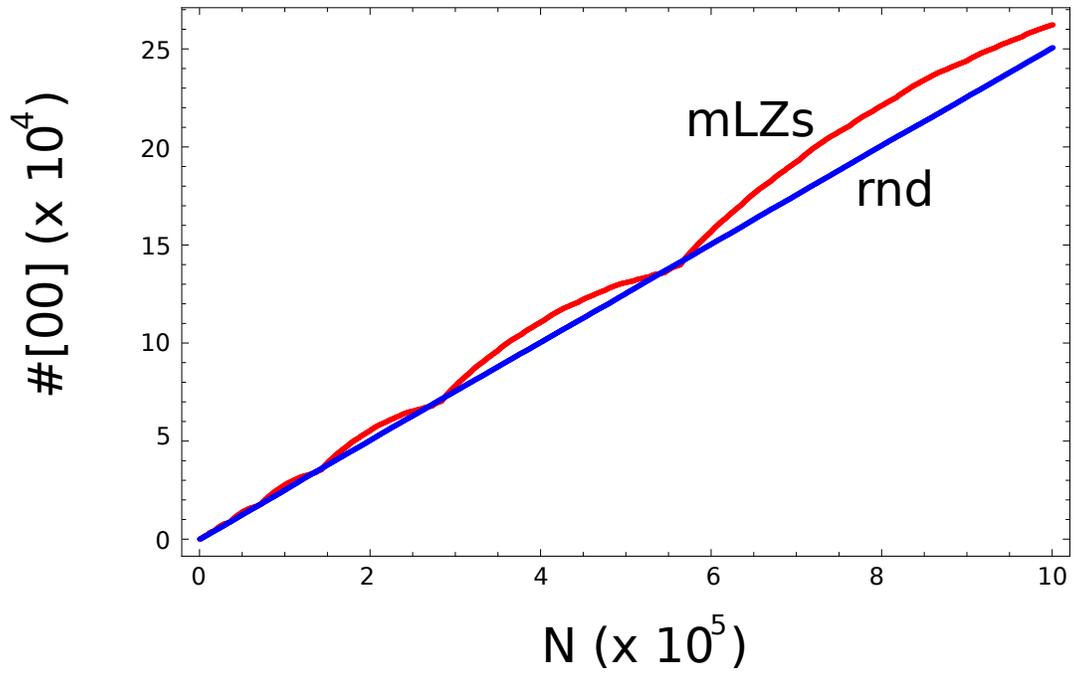}
\caption{Number of $00$ patterns ($\#[00]$) in the MLZs and random (rnd) sequences as a function of sequence length $N$.  While the $\#[00]$ for the random sequences exhibit the expected linear behavior with slope $1/4$, the behavior for the MLZs departs from a linear law.}\label{patt}
\end{figure}

\begin{figure}
\includegraphics[scale=1.5]{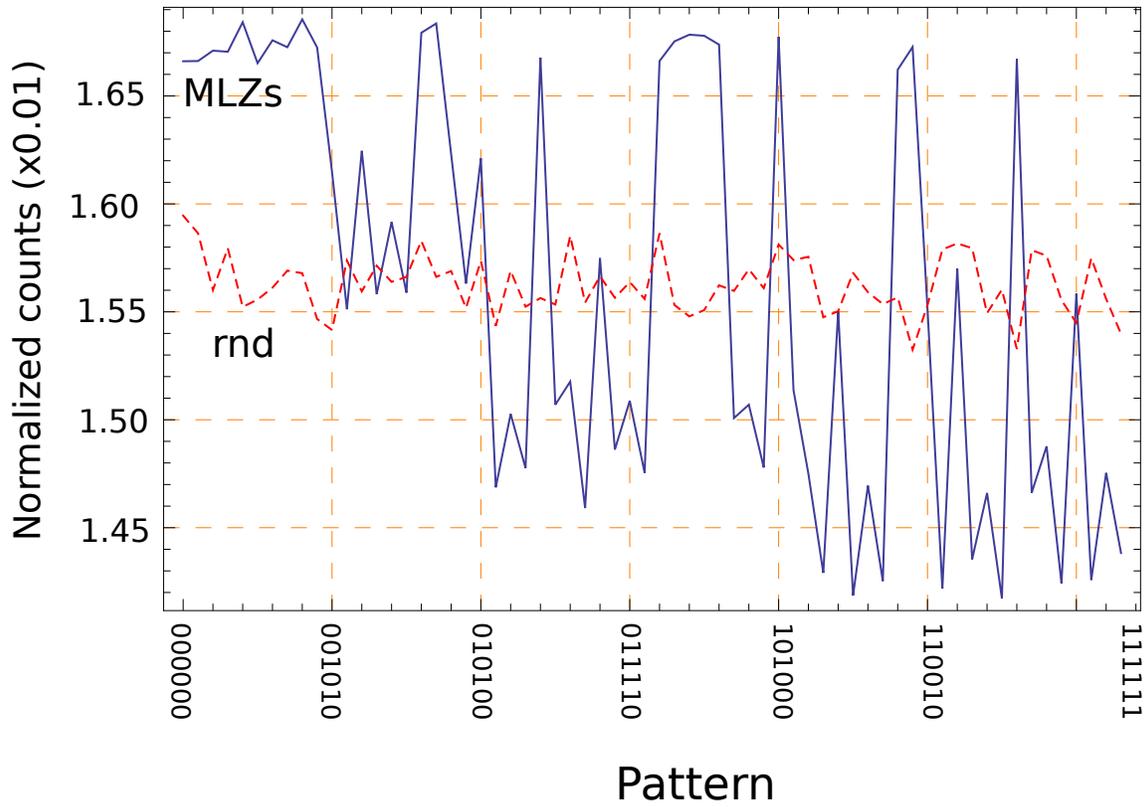}
\caption{Normalized counts for all patterns of length 6 in the MLZs and random (rnd) sequences (String length N=$10^6$). Patterns are ordered by their binary values. In the rnd curve, all patterns have counts near the expected value $2^{-6}(=0.0156)$, while for the MLZs, counts vary from slightly below 0.0141 to slightly above 0.0168.
}\label{allpatt}
\end{figure}

The MLZs are by construction non-stationary and non-random. For a binary source, in a $10^6$ length MLZs, the probability of occurrence of one of the symbol, is almost equal to the probability of the other symbol, giving a Shannon entropy around $1$ bit/symbol (The calculation was performed $5\times 10^3$ times with random ordering in the test step). It has been proved\cite{li94} that a finite random sequence with maximum KC complexity does not have the same number of zeros and ones. Furthermore, figure \ref{patt} shows the number of $00$ patterns of the MLZs compared to the random string for increasing length. The MLZs follow a different behavior than the random sequence. Normality of the random string, results in a linear behavior of the number of $00$ patterns, with slope $2^{-2}$. The number of $00$ patterns for the MLZs are mostly above the random curve, showing a multiple ''bumped'' curve. The non-random nature is further emphasized by figure \ref{allpatt}, where the normalized counts of all six length patterns in the MLZs is shown compared with the random sequence \cite{Note3}. Normal behavior, in the sense of Borel, can be observed for the random sequence (the probability of occurrence of all k-length patterns are equal, and tend to $\sigma^{-k}$ for infinite strings)  with a probability of occurrence near $2^{-6}(=0.0156)$. This is clearly not the case for the MLZs, which show different probability of occurrence between different substrings.

\begin{figure}
\includegraphics{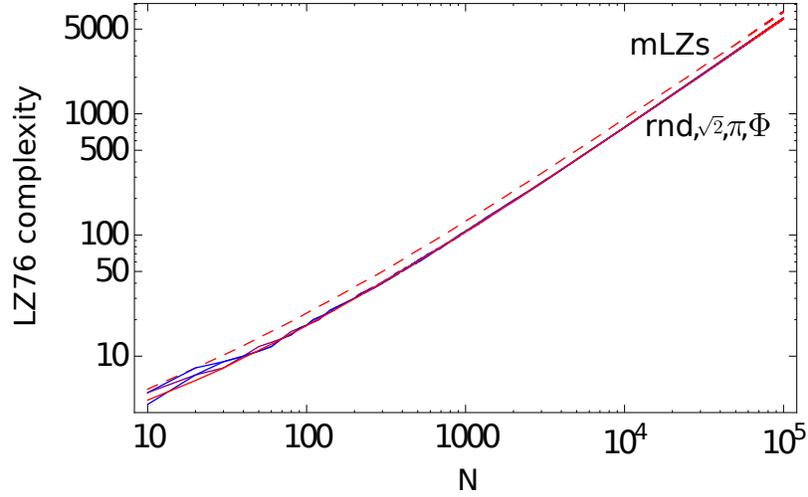}
\caption{LZ76 complexity as a function of the sequence length N (log-log scale) for the binary expansion of $\pi$,
$\sqrt{2}$ and $\Phi$ sequences, together with the MLZs and random (rnd) sequences. The binary expanded irrational numbers can not be distinguished from the random LZ76 complexity for all lengths considered. All LZ76 complexities are below the MLZs complexity.
}\label{irrationallz76}
\end{figure}

The fact that typical random sequences fall below the maximum attainable LZ76 complexity for a fixed finite length, and that MLZs are far from being random in any sense, suggest that it is possible that other non-random strings could achieve LZ76 complexities comparable to that of the random string. This is certainly the case, we compared the LZ76 complexity of the MLZs with that of the binary expansion of the irrational  numbers $\pi$,
$\sqrt{2}$ and $\Phi$ (the Golden Ratio). The last three sequences are thought to be normal in Borel sense\cite{wagon85}. Figure \ref{irrationallz76} shows the LZ76 complexity of the described sequences, together with that of the MLZs and the random sequence. LZ76 complexity of  $\pi$, $\sqrt{2}$ and $\Phi$ are indistinguishable from the random sequence; above a sequence length of $10^3$, all four sequences exhibit the same increasing law.

As already stated, Borel normal numbers are those numbers whose expansion in any base, have all possible patterns of a given length occurring with equal probability. It has been argued that, for example, the number $\pi$ behaves randomly in the frequency of occurrence of its digits in base ten up to the first ten million digits\cite{wagon85}. This fact also extends to the frequency of considering all digit pairs, digit triples, etc. The number $\pi$ seems to be normal, and indeed it has passed randomness statistical test\cite{arndt01}. If $\pi$ is normal, then any substring, including the initial digits $31415\dots$, will happen equally probable to any other substring of the same length. It is then interesting to ask if $\pi$ (or any other of the transcendental numbers considered) can be considered in some sense stationary. Even if this is the case, $\pi$ is certainly non random, as any arbitrary sequence  of digits can be effectively calculated\cite{arndt01} and therefore, its  KC complexity rate will be zero. Even more,  LZ76 is unable to discriminate a true random sequence from a perfectly algorithmically determined sequence, as that of the binary expansion of computable irrational numbers. The model based character of LZ76 complexity exhibits its narrower scope compared to the KC complexity.

\section{Normalization of Lempel-Ziv complexity}\label{sec:norm}

In order to compare different complexity measures of a sequence, some kind of normalization is needed. One approach would be to normalize by the LZ76 complexity of a random sequence of the same length. The problem with this normalization is that, as already seen, the LZ76 complexity of a random sequence is not an upper bound for the LZ76 complexity of the set of sequences of a given length. This results in a normalized complexity that does not lie in a definite interval, and normalized complexity above one can be found in practical measurements\cite{hu06}. We have also, already discussed that $c[u(1,N)]$ can have values above $1$ for finite size sequences.

Instead, the construction of the MLZs motivates an alternative normalization of the LZ76 complexity,

\begin{equation}\label{cmlzh}
 c_{mlz}(u)=\frac{C(u)}{C(MLZs)}.
\end{equation}

As $C(MLZs)$ is an upper bound, regardless of the sequence length, $c_{mlz}(u)$ will always be in the interval $[0,1]$. Furthermore, as the MLZs tend, by definition, to the asymptotic value of $\log N/N$, $c_{mlz}(u)$ will comply with Ziv theorem given by equation (\ref{zivtheorem}).

In reference [\onlinecite{schurmann99}] an empirical ansatz for estimating the entropy rates from empirical data, using estimates of different natures, has been reported,

\begin{equation}\label{ansatz}
 c(u)=h+b\frac{\log{N}}{N^c} \;\;\;\;\;\; c> 0,
\end{equation}

that yields excellent fits to sequences generated by different sources, and gives a good estimate of the entropy rate $h$ in all cases considered. The same normalization  has been further studied in reference [\onlinecite{lesne09}] for short sequences of around $10^3$ symbols, observing a good compromise to the estimated entropy rate. 

We compared the entropy rate estimation using the asymptotic value normalized LZ76 complexity $c(u)$, appearing in equation (\ref{zivtheorem}), with $c_{mlz}(u)$, given by equation (\ref{cmlzh}), for different sequence sources. Good fitting of the complexity function $c_{mlz}(u)$ was found with the following ansatz:

\begin{equation}\label{ansatzmlz}
 c_{mlz}(u)=h+a\varepsilon_{N}\log{N}+b\frac{\log{N}}{N^c} \;\;\;\;\;\;\;\;\; c > 0
\end{equation}

this empirical function was found to be superior than equation (\ref{ansatz}) in several cases.

\section{LZ76 complexity and complexity rate estimation for different sources}\label{sec:comparison}

Our first example is given by the logistic map

\begin{equation}\label{eq:lmap}
 x_{n+1}=1-r x_{n}^2 
\end{equation}

using a binary generating partition at $x=0$. This process was studied in the context of Lempel-Ziv complexity in reference [\onlinecite{schurmann99}]. Three cases with different parameter $r$ are of interest: the chaotic range $r=1.8$, with an entropy rate of $h=0.5828$; $r=1.7499$, where a strong intermittent point is observed, and an entropy rate $h=0.2597$ is achieved; and the Feigenbaum point at $r=1.40115518$ where the entropy rate becomes zero.

\begin{figure}
\includegraphics{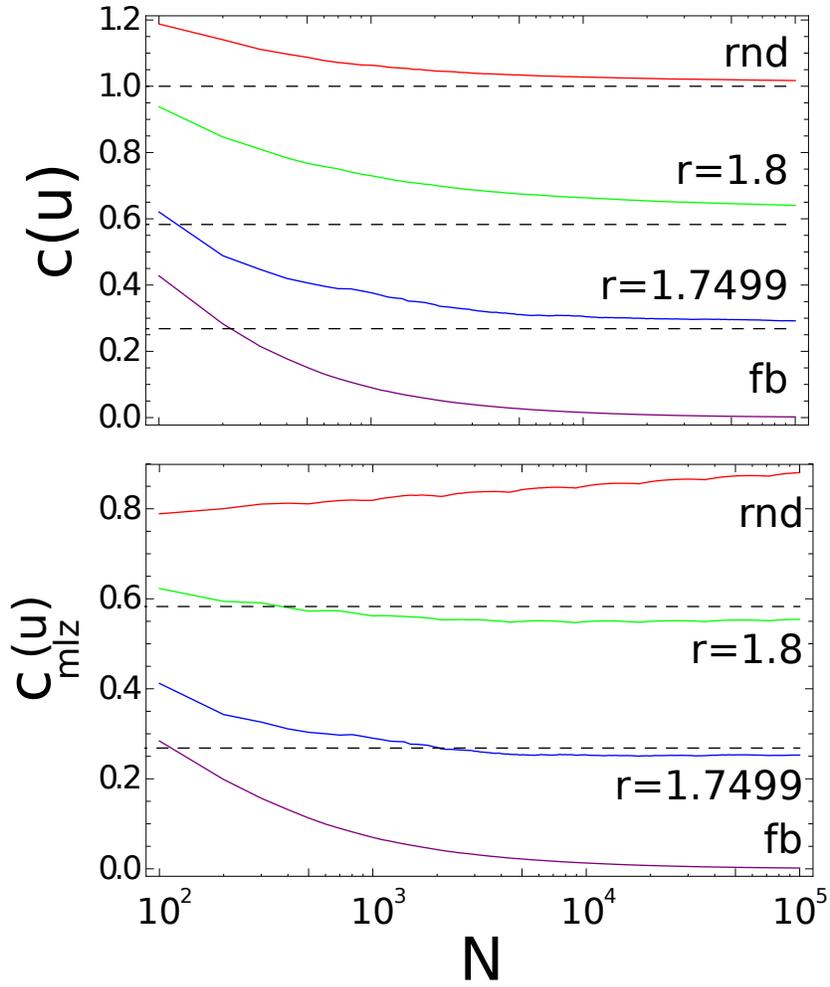}
\caption{$c(u)$ and $c_{mlz}(u)$ for the logistic map given by equation (\ref{eq:lmap}), and compared to the random (rnd) sequence. Three values for the logistic map parameter were considered: the chaotic regime $r=1.8$; the intermittent point $r=1.7499$; and the Feigenbaum point (fb) at $r=1.40115518$. See text for details.}\label{logmap}
\end{figure}

Figure \ref{logmap} shows the value of $c(u)$ and $c_{mlz}(u)$ for the three values of the $r$ parameter, together with the random sequence up to $N=10^5$ symbols. For each $r$ parameter, $50$ sequences were simulated and the LZ76 complexity value was averaged over the set. While $c(u)$ approximates the entropy rate value from above, $c_{mlz}(u)$ tends also to the entropy rate but from below. Table \ref{table:logmap} shows the entropy rate estimated as the value of $c(u)$ and $c_{mlz}(u)$ for the maximum string length. $c_{mlz}(u)$ compares better than $c(u)$ for all considered $r$ values, being the largest relative error of $4.9\%$ for the former estimate, less than two times compared to the best estimate for $c(u)$. It is also interesting to note that $c(u)$ behaves worse for the $r$ value for which $c_{mlz}(u)$ behaves better.

\begin{figure}
\includegraphics{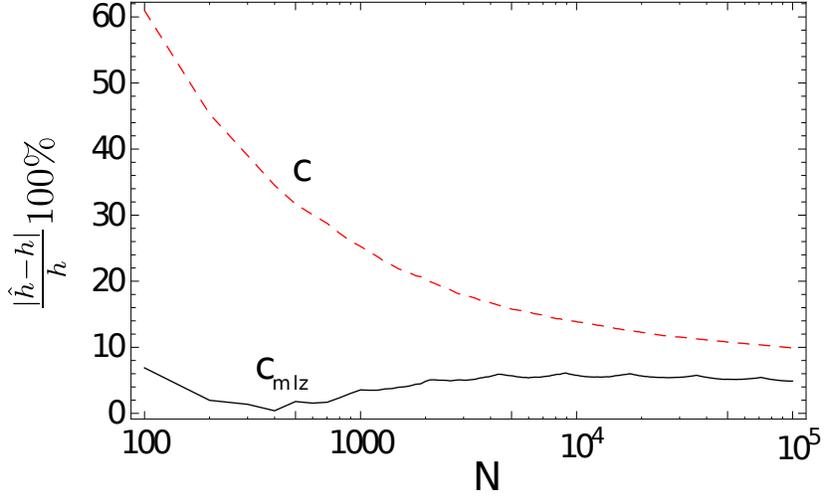}
\caption{The relative error of $c(u)$ and $c_{mlz}(u)$ estimates ($\hat{h}$) of the true entropy ($h$) for the $r=1.8$ logistic map as function of the sequence length N.}\label{relerror}
\end{figure}

Figure \ref{relerror} shows that $c_{mlz}(u)$ outperforms $c(u)$ for the whole range of sequence lengths.

We made a fit to the ansatz given by equations (\ref{ansatz}) and (\ref{ansatzmlz}), for $c(u)$ and $c_{mlz}(u)$, respectively. Table \ref{table:logmap} shows the results. $c(u)$ improves slightly the entropy rate estimates, while $c_{mlz}(u)$ improves dramatically for $r=1.8$, but a worse estimate of $h$ is attained for $r=1.7499$. All fits had a $R^2$ figure of merit above $0.999$.

We now consider as a second example, random sequences resulting from a $(\alpha, 1-\alpha)$-Bernoulli process (a biased coin toss). The Shannon
entropy rate is given by

\begin{equation}\label{eq:hrnd}
 h=-\alpha \log{\alpha}-(1-\alpha)\log{(1-\alpha)}
\end{equation}

The cases of $\alpha=1/2, 1/4, 1/16, 1/64$ where analyzed. Table \ref{table:rnd} gives the estimates of the entropy rate for $c(u)$ and $c_{mlz}(u)$ following the same procedure than that for the logistic map. Again $10^5$ characters were considered and for each $\alpha$ value, $50$ sequences were generated and the average complexity value was calculated.

In general, $c(u)$ performs better than $c_{mlz}(u)$, yet for the later, a significant improvement is achieved through the fit by equation (\ref{ansatzmlz}). For the later case, relative errors drops below $2\%$ in all cases except for the almost constant sequence $\alpha=1/64$, where the relative error is larger in all estimates, due to the low value of the entropy rate.

\begin{figure}
\includegraphics{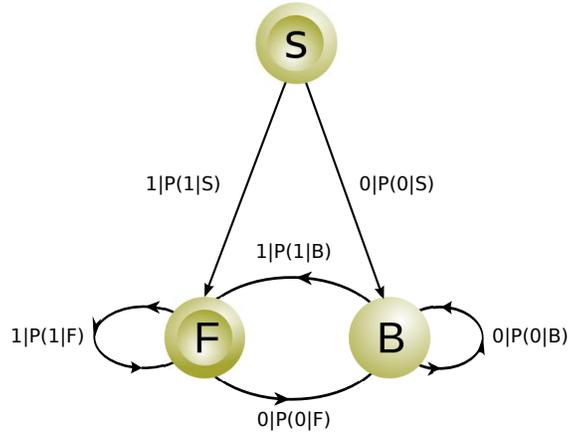}
\caption{Finite State Automata for the nearest neighbor interaction range. $S$ is the start state while $F$ and $B$, are recurrent states. $P(X|M)$ represents the probability of emitting a symbol $X$ conditioned on being in state $M$.}\label{fsa}
\end{figure}

Finally we considered the $2$-state automata depicted in figure \ref{fsa}. The directed arcs represent transitions between states with conditioned probability $P(X|M)$, where $X$ stands for the symbol being emitted, and $M$ is the current state. The $S$ state is the starting state which is transient, $F$ and $C$ are recurrent states. This finite state automata has been studied in the context of $\varepsilon$-machine, to describe the spin-$1/2$, nearest-neighbor Ising model. The reader is referred to reference [\onlinecite{feldman98}] for a detailed discussion of its statistical properties. For the purpose of this paper, it will suffice to know that the entropy rate of the stationary regime is given by

\begin{widetext}
\begin{equation}\label{hising}
 h=\frac{P(1|F)-1}{1-P(1|F)+P(1|B)}\sum_{i=0,1}P(i|B)-\frac{P(1|B)}{1-P(1|F)+P(1|B)}\sum_{i=0,1}P(i|F)
\end{equation}
\end{widetext}

and the normalization conditions

\begin{eqnarray*}
 P(1|B)+P(0|B)=1 \\
 P(1|F)+P(0|F)=1.
\end{eqnarray*}

Table \ref{table:ising} shows the estimates of the entropy rate $c(u)$, $c_{mlz}(u)$, as well as the fitted values for $10^5$ length sequences. For each length, $50$ sequences were generated and values were averaged. The third column gives the $h$ value calculated from equation (\ref{hising}).

Three cases were considered, (a) is within the ferromagnetic regime, while (b) is antiferromagnetic and, (c) is almost paramagnetic. Excellent agreement of the estimated entropies are attained for all cases both using $c(u)$ and $c_{mlz}(u)$. The performance of the entropy rate fitted values, to $c(u)$ and $c_{mlz}(u)$, are similar, with relative errors at most $2\%$.

\section{Conclusions}

The fact that MLZs are deterministic in nature, and therefore of vanishing Kolmogorov-Chaitin entropy rate, seems to go against the common idea that a typical random sequence of finite length N, attains maximum LZ76 complexity. Yet, the numerical simulations carried out, shows that typical random sequence, in a wide range of length values, falls  short of maximum complexity. Furthermore, the idea that LZ76 complexity is higher, the higher the randomness of a sequence, proves also to be wrong in the finite length case. This does not contradict the theorem that in the infinite limit, the LZ76 complexity rate is closely related to the entropy rate for an ergodic  source. 

The analysis of the binary expansion of certain irrational numbers, showed a LZ76 complexity behavior, as a function of the sequence length, indistinguishable from a random sequence. In some sense, this comes as no surprise, in spite of having a zero KC complexity rate, numbers such as $\pi$ are known to pass random test of different sort. LZ76 complexity is also ''fooled'' by the nature of such sequences. Grassberger\cite{grassberger} has justified the  random appearance of $\pi$ with the argument, that while complexity estimates, like LZ76, measure an information related quantity characteristic of the first $N$ digits of a sequence, entropy rate $h$, as any statistical information quantity, measures an ensemble average of the source output that has to do with any substring of length $N$. Compelling as it may seems, one might wonder, assuming Borel normality, if some sort of stationary behavior must be expected for any finite size substring, which will render such argument as weak. In any case, it is clear that while KC complexity can not be ''tricked'' by an intelligent agent, LZ76 complexity measure has a much narrower scope due to its model based nature.

Finally, simulations show that the normalization of LZ76 complexity by the complexity of MLZs of the same length, can be used to find good estimates of the entropy rate, while keeping its value bounded in the interval $[0,1]$. An empirical ansatz was found with excellent fit to the studied data. While the examples examined in this paper can not exhaust all possibilities, it shows at least, that for a wide range of sources,  $c_{mlz}$ can be used to estimate the entropy rate in practical experimental data analysis. Even if normalization by the LZ76 complexity of  a random sequence is still preferred, comparison of normalized values with the LZ76 complexity of the MLZs still can prove useful in a number of cases.

\begin{acknowledgments}

E. Estevez-Rams which to thanks the Humboldt Foundation (AvH) for computational infrastructure support and the FAPEMIG
(BPV-00039-12) for financial assistance in mobility.
\end{acknowledgments}

\pagebreak
\pagebreak
\begin{table}
\begin{tabular}{*{6}{p{50pt}}}
 \toprule
r      & h             & $c(u)$               & fitted h             & $c_{mlz}(u)$  & fitted $h_{mlz}$           \\
\hline                                                            
1.8000 & 0.5828        & 0.6405		      &  0.6348              &  0.5544       & 0.5883                      \\
       &               & (9.9 \%)             &  (8,9 \%)            &  (4.9\%)      & (0.94\%)                    \\
\hline
1.7499 & 0.2597        & 0.2919               &  0.2891              &  0.2526       & 0.2882                      \\
       &               & (12.5\%)             &  (11.3 \%)           &  (2.7 \%)     & (11\%)                      \\
\hline
1.4011 & 0             & 0.0020               & 0.0020               & 0.0018        & 0.012                       \\
\hline 
\end{tabular}

\caption{Estimated entropies (bit/symbol) for the logistic map (Equation (\ref{eq:lmap})). The second column is the entropy rate value from reference (\onlinecite{schurmann99}). Third and fifth column are the value of $c(u)$ and $c_{mlz}(u)$, respectively, for the $10^5$ length sequence, each value is averaged over 50 sequences. Fourth and six column corresponds to the entropy rate estimated by fitting the values of $c(u)$ (equation \ref{ansatz}) and $c_{mlz}(u)$ (equation (\ref{ansatzmlz})), respectively. Values between round brackets are the relative errors with respect to columns two values.}
\label{table:logmap}
\end{table}

\begin{table}
\begin{tabular}{*{6}{p{50pt}}}
 \toprule
$\alpha$      & h             & $c(u)$               & fitted h             & $c_{mlz}(u)$  & fitted $h_{mlz}$ \\
\hline
$1/2$         & 1             & 1.017                &  1.013               &  0.88        & 1.019             \\
              &               & (1.7\%)              &  (1.4 \%)            &  (12 \%)     & (1.9\%)           \\
\hline
$1/4$         & 0.8113        & 0.8144               & 0.8129               & 0.704        & 0.8243             \\
              &               & (0.4\%)              &  (0.2 \%)            &  (13 \%)     & (1.6\%)             \\
\hline
$1/16$        & 0.3373        & 0.3233               & 0.3221               & 0.2798        & 0.3370             \\
              &               & (4.1\%)              &  (4.5 \%)            &  (17 \%)      & (0.09\%)             \\
\hline
$1/64$        & 0.1161        & 0.1062               & 0.1060               & 0.0919        & 0.0981              \\
              &               & (8.5\%)              &  (8.6 \%)            &  (20.8\%)     & (15\%)             \\
\hline
\end{tabular}

\caption{Estimated entropy (bit/symbol) for a $(\alpha,1-\alpha)$-Bernoulli process. See Table \ref{table:logmap}
for a description of the column values.}
\label{table:rnd}
\end{table}

\begin{table}
\begin{tabular}{*{8}{p{50pt}}}
 \toprule
  & $P(1|F)$ & $P(1|B)$ & h      & $c(u)$    & fitted h      & $c_{mlz}(u)$  & fitted $h_{mlz}$ \\
\hline
a & 0.90     & 0.39     & 0.558  &  0.564    &  0.564        & 0.488         & 0.550             \\
  &          &          &        &           & (1.1 \%)      &               & (1.4\%)           \\
\hline
b & 0.56     & 0.95     & 0.767  &  0.783    & 0.781         & 0.677         & 0.779              \\
  &          &          &        &           & (1.8 \%)      &               & (1.6\%)            \\
\hline
c & 0.53     & 0.46     & 0.996  & 1.01      & 1.010         & 0.877         & 1.017              \\
  &          &          &        &           & (1.4 \%)      &               & (2.1\%)            \\
\hline
\end{tabular}

\caption{Estimated entropy (bit/symbol) for nearest neighbor Ising model. $P(X|M)$ represents the probability of emitting a symbol $X$ conditioned by being on state $M$. The rest of the columns follows the same description than Table \ref{table:logmap}.}
\label{table:ising}
\end{table}

\begin{thebibliography}{10}%
\makeatletter
\providecommand \@ifxundefined [1]{%
 \ifx #1\undefined \expandafter \@firstoftwo
 \else \expandafter \@secondoftwo
\fi
}%
\providecommand \@ifnum [1]{%
 \ifnum #1\expandafter \@firstoftwo
 \else \expandafter \@secondoftwo
\fi
}%
\providecommand \enquote [1]{``#1''}%
\providecommand \bibnamefont  [1]{#1}%
\providecommand \bibfnamefont [1]{#1}%
\providecommand \citenamefont [1]{#1}%
\providecommand\href[0]{\@sanitize\@href}%
\providecommand\@href[1]{\endgroup\@@startlink{#1}\endgroup\@@href}%
\providecommand\@@href[1]{#1\@@endlink}%
\providecommand \@sanitize [0]{\begingroup\catcode`\&12\catcode`\#12\relax}%
\@ifxundefined \pdfoutput {\@firstoftwo}{%
 \@ifnum{\z@=\pdfoutput}{\@firstoftwo}{\@secondoftwo}%
}{%
 \providecommand\@@startlink[1]{\leavevmode}%
 \providecommand\@@endlink[0]{}%
}{%
 \providecommand\@@startlink[1]{%
  \leavevmode
  \pdfstartlink
   attr{/Border[0 0 1 ]/H/I/C[0 1 1]}%
   user{/Subtype/Link/A<</Type/Action/S/URI/URI(#1)>>}%
  \relax
 }%
 \providecommand\@@endlink[0]{\pdfendlink}%
}%
\providecommand \url  [0]{\begingroup\@sanitize \@url }%
\providecommand \@url [1]{\endgroup\@href {#1}{\urlprefix}}%
\providecommand \urlprefix [0]{URL }%
\providecommand \Eprint[0]{\href }%
\@ifxundefined \urlstyle {%
  \providecommand \doi [1]{doi:\discretionary{}{}{}#1}%
}{%
  \providecommand \doi [0]{doi:\discretionary{}{}{}\begingroup
  \urlstyle{rm}\Url }%
}%
\providecommand \doibase [0]{http://dx.doi.org/}%
\providecommand \Doi[1]{\href{\doibase#1}}%
\providecommand \selectlanguage [0]{\@gobble}%
\providecommand \bibinfo [0]{\@secondoftwo}%
\providecommand \bibfield [0]{\@secondoftwo}%
\providecommand \translation [1]{[#1]}%
\providecommand \BibitemOpen[0]{}%
\providecommand \bibitemStop [0]{}%
\providecommand \bibitemNoStop [0]{.\EOS\space}%
\providecommand \EOS [0]{\spacefactor3000\relax}%
\providecommand \BibitemShut [1]{\csname bibitem#1\endcsname}%
\bibitem{lz76}%
  \BibitemOpen
  \bibfield{author}{%
  \bibinfo {author} {\bibfnamefont{A.}~\bibnamefont{Lempel}}\ and\ \bibinfo
  {author} {\bibfnamefont{J.}~\bibnamefont{Ziv}},\ }%
  \bibfield{title}{%
  \enquote{\bibinfo {title} {On the complexity of finite sequences}.}\ }%
  \bibfield{journal}{%
  \bibinfo {journal} {IEEE Trans. Inf. Th.}\ }%
  \textbf{\bibinfo {volume} {IT-22}},\ \bibinfo {pages} {75--81} (\bibinfo
  {year} {1976})\BibitemShut{NoStop}%
\bibitem{amigo03}%
  \BibitemOpen
  \bibfield{author}{%
  \bibinfo {author} {\bibfnamefont{J.~M.}\ \bibnamefont{Amigo}}, \bibinfo
  {author} {\bibfnamefont{J.}~\bibnamefont{Szczepanski}}, \bibinfo {author}
  {\bibfnamefont{E.}~\bibnamefont{Wajnryb}},\ and\ \bibinfo {author}
  {\bibfnamefont{M.~V.}\ \bibnamefont{Sanchez-Vives}},\ }%
  \bibfield{title}{%
  \enquote{\bibinfo {title} {On the number of states of the neuronal
  sources},}\ }%
  \bibfield{journal}{%
  \bibinfo {journal} {Biosystems}\ }%
  \textbf{\bibinfo {volume} {68}},\ \bibinfo {pages} {57--66} (\bibinfo {year}
  {2003})\BibitemShut{NoStop}%
\bibitem{szczepanski04}%
  \BibitemOpen
  \bibfield{author}{%
  \bibinfo {author} {\bibfnamefont{J.}~\bibnamefont{Szczepanski}}, \bibinfo
  {author} {\bibfnamefont{J.~M.}\ \bibnamefont{Amigo}}, \bibinfo {author}
  {\bibfnamefont{E.}~\bibnamefont{Wajnryb}},\ and\ \bibinfo {author}
  {\bibfnamefont{M.~V.}\ \bibnamefont{Sanchez-Vives}},\ }%
  \bibfield{title}{%
  \enquote{\bibinfo {title} {Characterizing spike trains with Lempel-Ziv
  complexity},}\ }%
  \bibfield{journal}{%
  \bibinfo {journal} {Neurocomp.}\ }%
  \textbf{\bibinfo {volume} {58-60}},\ \bibinfo {pages} {77--84} (\bibinfo
  {year} {2004})\BibitemShut{NoStop}%
\bibitem{abo06}%
  \BibitemOpen
  \bibfield{author}{%
  \bibinfo {author} {\bibfnamefont{M.}~\bibnamefont{Aboy}}, \bibinfo {author}
  {\bibfnamefont{R.}~\bibnamefont{Homero}}, \bibinfo {author}
  {\bibfnamefont{D.}~\bibnamefont{Abasolo}},\ and\ \bibinfo {author}
  {\bibfnamefont{D.}~\bibnamefont{Alvarez}},\ }%
  \bibfield{title}{%
  \enquote{\bibinfo {title} {Interpretation of the lempel-ziv complexity
  measure in the context of biomedical signal analysis},}\ }%
  \bibfield{journal}{%
  \bibinfo {journal} {IEEE Trans. Biom. Eng.}\ }%
  \textbf{\bibinfo {volume} {53}},\ \bibinfo {pages} {2282--2288} (\bibinfo
  {year} {2006})\BibitemShut{NoStop}%
\bibitem{hu06}%
  \BibitemOpen
  \bibfield{author}{%
  \bibinfo {author} {\bibfnamefont{J.}~\bibnamefont{Hu}}, \bibinfo {author}
  {\bibfnamefont{J.}~\bibnamefont{Gao}},\ and\ \bibinfo {author}
  {\bibfnamefont{J.~C.}\ \bibnamefont{Principe}},\ }%
  \bibfield{title}{%
  \enquote{\bibinfo {title} {Analysis of biomedical signals by the Lempel-Ziv
  complexity: the effect of finite data size},}\ }%
  \bibfield{journal}{%
  \bibinfo {journal} {IEEE Trans. Biomed. Eng.}\ }%
  \textbf{\bibinfo {volume} {53}},\ \bibinfo {pages} {2606--2609} (\bibinfo
  {year} {2006})\BibitemShut{NoStop}%
\bibitem{zhang09}%
  \BibitemOpen
  \bibfield{author}{%
  \bibinfo {author} {\bibfnamefont{Y.}~\bibnamefont{Zhang}}, \bibinfo {author}
  {\bibfnamefont{J.}~\bibnamefont{Hao}}, \bibinfo {author}
  {\bibfnamefont{C.}~\bibnamefont{Zhou}},\ and\ \bibinfo {author}
  {\bibfnamefont{K.}~\bibnamefont{Chang}},\ }%
  \bibfield{title}{%
  \enquote{\bibinfo {title} {Normalized Lempel-Ziv complexity and its
  applications in biosequence analysis},}\ }%
  \bibfield{journal}{%
  \bibinfo {journal} {J. Math. Chem.}\ }%
  \textbf{\bibinfo {volume} {46}},\ \bibinfo {pages} {1203--1212} (\bibinfo
  {year} {2009})\BibitemShut{NoStop}%
\bibitem{ziv78}%
  \BibitemOpen
  \bibfield{author}{%
  \bibinfo {author} {\bibfnamefont{J.}~\bibnamefont{Ziv}},\ }%
  \bibfield{title}{%
  \enquote{\bibinfo {title} {Coding theorems for individual sequences},}\ }%
  \bibfield{journal}{%
  \bibinfo {journal} {IEEE Trans. Inf. Th.}\ }%
  \textbf{\bibinfo {volume} {IT-24}},\ \bibinfo {pages} {405--412} (\bibinfo
  {year} {1978})\BibitemShut{NoStop}%
\bibitem{chelani11}%
  \BibitemOpen
  \bibfield{author}{%
  \bibinfo {author} {\bibfnamefont{A.B.}\ \bibnamefont{Chelani}},\ }%
  \bibfield{title}{%
  \enquote{\bibinfo {title} {Complexity analysis of co concentrations at a
  trafﬁc site in Delhi},}\ }%
  \bibfield{journal}{%
  \bibinfo {journal} {Transp. Res. D}\ }%
  \textbf{\bibinfo {volume} {16}},\ \bibinfo {pages} {57--60} (\bibinfo {year}
  {2011})\BibitemShut{NoStop}%
\bibitem{rajkovic03}%
  \BibitemOpen
  \bibfield{author}{%
  \bibinfo {author} {\bibfnamefont{M.}~\bibnamefont{Rajkovic}}\ and\ \bibinfo
  {author} {\bibfnamefont{Z.}~\bibnamefont{Mihailovic}},\ }%
  \bibfield{title}{%
  \enquote{\bibinfo {title} {Quantifying complexity in the minority game},}\ }%
  \bibfield{journal}{%
  \bibinfo {journal} {Physica A}\ }%
  \textbf{\bibinfo {volume} {325}},\ \bibinfo {pages} {40--47} (\bibinfo {year}
  {2003})\BibitemShut{NoStop}%
\bibitem{liu12}%
  \BibitemOpen
  \bibfield{author}{%
  \bibinfo {author} {\bibfnamefont{L.}~\bibnamefont{Liu}}, \bibinfo {author}
  {\bibfnamefont{D.}~\bibnamefont{Li}},\ and\ \bibinfo {author}
  {\bibfnamefont{F.}~\bibnamefont{Bai}},\ }%
  \bibfield{title}{%
  \enquote{\bibinfo {title} {A relative Lempel–Ziv complexity: Application to
  comparing biological sequences},}\ }%
  \bibfield{journal}{%
  \bibinfo {journal} {Chem. Phys. Lett.}\ }%
  \textbf{\bibinfo {volume} {530}},\ \bibinfo {pages} {107--112} (\bibinfo
  {year} {2012})\BibitemShut{NoStop}%
\bibitem{rapp01}%
  \BibitemOpen
  \bibfield{author}{%
  \bibinfo {author} {\bibfnamefont{P.~E.}\ \bibnamefont{Rapp}}, \bibinfo
  {author} {\bibfnamefont{C.~J.}\ \bibnamefont{Cellucci}}, \bibinfo {author}
  {\bibfnamefont{K.~E.}\ \bibnamefont{Korslund}}, \bibinfo {author}
  {\bibfnamefont{T.~A.~A.}\ \bibnamefont{Watanabe}},\ and\ \bibinfo {author}
  {\bibfnamefont{M.~A.}\ \bibnamefont{Jimenez-Montano}},\ }%
  \bibfield{title}{%
  \enquote{\bibinfo {title} {Effective normalization of complexity measurements
  for epoch length and sampling frequency},}\ }%
  \bibfield{journal}{%
  \bibinfo {journal} {Phys. Rev. E}\ }%
  \textbf{\bibinfo {volume} {64}},\ \bibinfo {pages} {016209--016217} (\bibinfo
  {year} {2001})\BibitemShut{NoStop}%
\bibitem{kolmogorov65}%
  \BibitemOpen
  \bibfield{author}{%
  \bibinfo {author} {\bibfnamefont{A.~N.}\ \bibnamefont{Kolmogorov}},\ }%
  \bibfield{title}{%
  \enquote{\bibinfo {title} {Three approaches to the concept of the amount of
  information.}.}\ }%
  \bibfield{journal}{%
  \bibinfo {journal} {Probl. Inf. Transm. (English Trans.).}\ }%
  \textbf{\bibinfo {volume} {1}},\ \bibinfo {pages} {1--7} (\bibinfo {year}
  {1965})\BibitemShut{NoStop}%
\bibitem{martin66}%
  \BibitemOpen
  \bibfield{author}{%
  \bibinfo {author} {\bibfnamefont{P.}~\bibnamefont{Martin-Lof}},\ }%
  \bibfield{title}{%
  \enquote{\bibinfo {title} {On the definition of random sequences},}\ }%
  \bibfield{journal}{%
  \bibinfo {journal} {Inf. and Control}\ }%
  \textbf{\bibinfo {volume} {9}},\ \bibinfo {pages} {602--619} (\bibinfo {year}
  {1966})\BibitemShut{NoStop}%
\bibitem{li94}%
  \BibitemOpen
  \bibfield{author}{%
  \bibinfo {author} {\bibfnamefont{M.}~\bibnamefont{Li}}\ and\ \bibinfo
  {author} {\bibfnamefont{P.~M.~B.}\ \bibnamefont{Vitanyi}},\ }%
  \bibfield{title}{%
  \enquote{\bibinfo {title} {Statistical properties of finite sequences with
  high Kolmogorov complexity},}\ }%
  \bibfield{journal}{%
  \bibinfo {journal} {Math. Sys. Th.}\ }%
  \textbf{\bibinfo {volume} {27}},\ \bibinfo {pages} {365--376} (\bibinfo
  {year} {1994})\BibitemShut{NoStop}%
\bibitem{calude02}%
  \BibitemOpen
  \bibfield{author}{%
  \bibinfo {author} {\bibfnamefont{C.~S.}\ \bibnamefont{Calude}},\ }%
  \emph{\bibinfo {title} {Information and Randomness.}}\ (\bibinfo {publisher}
  {Springer Verlag},\ \bibinfo {year} {2002})\BibitemShut{NoStop}%
\bibitem{Note1}%
  \BibitemOpen
  \bibinfo {note} {Typical string will be taken in the sense defined by
  information theory, see for example reference \protect \rev@citealpnum
  {cover06}}\BibitemShut{NoStop}%
\bibitem{lesne09}%
  \BibitemOpen
  \bibfield{author}{%
  \bibinfo {author} {\bibfnamefont{A.}~\bibnamefont{Lesne}}, \bibinfo {author}
  {\bibnamefont{J.L.Blanc}},\ and\ \bibinfo {author}
  {\bibfnamefont{L.}~\bibnamefont{Pezard}},\ }%
  \bibfield{title}{%
  \enquote{\bibinfo {title} {Entropy estimation of very short symbolic
  sequences}.}\ }%
  \bibfield{journal}{%
  \bibinfo {journal} {Phys. Rev. E}\ }%
  \textbf{\bibinfo {volume} {79}},\ \bibinfo {pages} {046208--046217} (\bibinfo
  {year} {2009})\BibitemShut{NoStop}%
\bibitem{Notereferee}%
  \BibitemOpen
  \bibinfo {note} {Almost surely is taken to mean with probability one, as was correctly pointed out by one referee}\BibitemShut{NoStop}%
\bibitem{Note2}%
  \BibitemOpen
  \bibinfo {note} {Also known as metric entropy, or Kolmogorov-Sinai
  entropy}\BibitemShut{NoStop}%
\bibitem{cover06}%
  \BibitemOpen
  \bibfield{author}{%
  \bibinfo {author} {\bibfnamefont{T.~M.}\ \bibnamefont{Cover}}\ and\ \bibinfo
  {author} {\bibfnamefont{J.~A.}\ \bibnamefont{Thomas}},\ }%
  \emph{\bibinfo {title} {Elements of information theory. Second edition}}\
  (\bibinfo {publisher} {Wiley Interscience, New Jersey},\ \bibinfo {year}
  {2006})\BibitemShut{NoStop}%
\bibitem{lz77}%
  \BibitemOpen
  \bibfield{author}{%
  \bibinfo {author} {\bibfnamefont{J.}~\bibnamefont{Ziv}}\ and\ \bibinfo
  {author} {\bibfnamefont{A.}~\bibnamefont{Lempel}},\ }%
  \bibfield{title}{%
  \enquote{\bibinfo {title} {A universal algorithm for sequential data-compression}.}\ }%
  \bibfield{journal}{%
  \bibinfo {journal} {IEEE Trans. Inf. Th.}\ }%
  \textbf{\bibinfo {volume} {IT-23}},\ \bibinfo {pages} {337--343} (\bibinfo
  {year} {1977})\BibitemShut{NoStop}%
\bibitem{Note4}%
  \BibitemOpen
  \bibinfo {note} {Although used, entropy estimates based on compression algorithms and software such as the UNIX utility gzip, or the family of pkzip software are misleading if not plainly wrong, as simple numeric examples can prove}\BibitemShut{NoStop}%
\bibitem{schurmann99}%
  \BibitemOpen
  \bibfield{author}{%
  \bibinfo {author} {\bibfnamefont{T.}~\bibnamefont{Schurmann}}\ and\ \bibinfo
  {author} {\bibfnamefont{P.}~\bibnamefont{Grassberg}},\ }%
  \bibfield{title}{%
  \enquote{\bibinfo {title} {Entropy estimation of symbol sequence.}.}\ }%
  \bibfield{journal}{%
  \bibinfo {journal} {Chaos}\ }%
  \textbf{\bibinfo {volume} {6}},\ \bibinfo {pages} {414--427} (\bibinfo {year}
  {1999})\BibitemShut{NoStop}%
\bibitem{Note3}%
  \BibitemOpen
  \bibinfo {note} {Substrings of length six are well below the effective length
  for a $10^6$ length string with $h=1$. The effective length gives an upper
  limit above which, statistical fluctuations due to the the finite size
  character of the sequence, starts to affect calculation. See reference
  [\protect \rev@citealpnum {lesne09}] for further discussion of effective
  length in the entropy analysis of finite sequences.}\BibitemShut{Stop}%
\bibitem{wagon85}%
  \BibitemOpen
  \bibfield{author}{%
  \bibinfo {author} {\bibfnamefont{S.}~\bibnamefont{Wagon}},\ }%
  \bibfield{title}{%
  \enquote{\bibinfo {title} {Is $\pi$ normal ?}.}\ }%
  \bibfield{journal}{%
  \bibinfo {journal} {Mathem. Intell.}\ }%
  \textbf{\bibinfo {volume} {7}},\ \bibinfo {pages} {65--67} (\bibinfo {year}
  {1985})\BibitemShut{NoStop}%
\bibitem{arndt01}%
  \BibitemOpen
  \bibfield{author}{%
  \bibinfo {author} {\bibfnamefont{J.}~\bibnamefont{Arndt}}\ and\ \bibinfo
  {author} {\bibfnamefont{C.}~\bibnamefont{Haenel}},\ }%
  \emph{\bibinfo {title} {$\pi$ unleashed}}\ (\bibinfo {publisher} {Springer
  Verlag},\ \bibinfo {year} {2001})\BibitemShut{NoStop}%
\bibitem{feldman98}%
  \BibitemOpen
  \bibfield{author}{%
  \bibinfo {author} {\bibfnamefont{D.~P.}\ \bibnamefont{Feldman}},\ }%
  \enquote{\bibinfo {title} {Computational mechanics of classical spin
  systems.}.}\  (\bibinfo {year} {1998}),\
  \Eprint{http://arxiv.org/abs/http://hornacek.coa.edu/dave/Thesis/thesis.html%
}{http://hornacek.coa.edu/dave/Thesis/thesis.html}\BibitemShut{NoStop}%
\bibitem{grassberger}%
  \BibitemOpen
  \bibfield{author}{%
  \bibinfo {author} {\bibnamefont{P.Grassberger}},\ }%
  \bibfield{title}{%
  \enquote{\bibinfo {title} {{Randomness, Information, and Complexity}},}\ }%
  \bibfield{journal}{%
  \bibinfo {journal} {ArXiv e-prints}}%
   (\bibinfo {year} {2012}),\
  \Eprint{http://arxiv.org/abs/1208.3459}{arXiv:1208.3459
  [physics.data-an]}\BibitemShut{NoStop}%
\end{thebibliography}
\end{document}